\documentclass[12pt, a4paper]{amsart}

\usepackage{amsfonts, amsmath, amssymb,latexsym}
\usepackage[english]{babel}
\usepackage[utf8]{inputenc}
\usepackage{url}
\usepackage[top=0.8in, left=1in, right=1in, bottom=1in]{geometry}

\usepackage{mathptmx}
\usepackage{times}

\def\R{{\Bbb R}}
\def\C{{\Bbb C}}
\def\F{{\Bbb F}}

\def\cl{{C}\!\ell}

\def\T{{\rm T}}
\def\SU{{\rm SU}}
\def\diag{{\rm diag}}

\def\SU{{\rm SU}}
\def\su{{\rm su}}

\def\Mat{{\rm Mat}}
\def\tr{{\rm tr}}
\newtheorem{thm}{Theorem}[section]
\def\{{\lbrace}
\def\}{\rbrace}
\def\st{\stackrel}

\begin{document}

\pagestyle{empty}
\title[Constant solutions of Yang-Mills equations and generalized Proca equations]
{Constant solutions of Yang-Mills equations\\ and generalized Proca equations}

\thanks{The reported study was funded by RFBR according to the research project No. 16-31-00347 mol\_a.}%

\maketitle

\vspace{-24pt}

\begin{center}
\author{{\bf N.~G.~Marchuk},\,$^a$ \,{\bf D.~S.~Shirokov}\,$^b$ $^c$}%

\small
$^a$ Steklov Mathematical Institute,\\ Russian Academy of Sciences, Moscow, Russia,\\ nmarchuk@mi.ras.ru

\vspace{6pt}

$^b$ Department of Mathematics, Faculty of Economic Sciences,\\ National Research University Higher School of Economics, Moscow, Russia,\\
dshirokov@hse.ru

\vspace{6pt}
$^c$ Kharkevich Institute for Information Transmission Problems,\\ Russian Academy of Sciences, Moscow, Russia,\\ shirokov@iitp.ru

\end{center}



\begin{abstract}
In this paper we present some new equations which we call Yang-Mills-Proca equations (or generalized Proca equations). This system of equations is a generalization of Proca equation and Yang-Mills equations and it is not gauge invariant. We present a number of constant solutions of this system of equations in the case of arbitrary Lie algebra. In details we consider the case when this Lie algebra is Clifford algebra or Grassmann algebra. We consider solutions of Yang-Mills equations in the form of perturbation theory series near the constant solution.
\end{abstract}

\thispagestyle{empty}

\section*{Introduction}

In this paper we present some new equations which we call Yang-Mills-Proca equations. This system of equations is a generalization of Proca equation and Yang-Mills equations. In Sections 1-3 we present some well-known facts about Maxwell's equations, Proca equation, and Yang-Mills equations and give citations to the literature.

A. Proca introduce his equation as a generalization of Maxwell's equations. It later emerged that Proca equation describes massive particle of spin 1. In a similar manner we generalize Yang-Mills equations. New equations can be considered as the partial case of Yang-Mills equations with some current that satisfies an additional condition. Yang-Mills theory provides models for all fundamental types of interactions.

The considered system of Yang-Mills-Proca equations is not gauge invariant, but it is invariant with respect to a global transformation which depends on some Lie group (see section 4). We are interested in constant solutions of this system of equations. We study the corresponding algebraic system of cubic equations in the case of arbitrary Lie algebra $L$ and present some solutions to this system (sections 5 and 6). In details we consider the case when this Lie algebra is Clifford algebra with respect to the commutator (see sections 8 and 9). In the section 7 we consider solutions of Yang-Mills equations in the form of perturbation theory series near constant solutions.

\section{Relativistic form of Maxwell's equations}

Let $\R^{1,3}$ be the Minkowski  space with Cartesian coordinates $x^\mu$, \ $\mu=0,1,2,3$ and let $\partial_\mu=\partial/{\partial x^\mu}$ be partial derivatives. The metric tensor of the Minkowski space is given by the diagonal matrix
\begin{equation}
\eta=\|\eta_{\mu\nu}\|=\|\eta^{\mu\nu}\|=\diag(1,-1,-1,-1).\label{eta13}
\end{equation}
Components of tensors (tensor fields) are enumerated by small Greek letters.
If we consider a tensor field of type $(r,s)$ and of rank $r+s$ with components $u^{\mu_1\ldots\mu_r}_{\nu_1\ldots\nu_s}=u^{\mu_1\ldots\mu_r}_{\nu_1\ldots\nu_s}(x)$, $x\in\R^{1,3}$, then we write
$u^{\mu_1\ldots\mu_r}_{\nu_1\ldots\nu_s}\in\T^r_s$ or $u\in\T^r_s$.
  With the aid of metric tensor we can raise or lower indices of components of tensor fields. For example, $f^{\mu\nu}=\eta^{\mu\alpha}\eta^{\nu\beta}f_{\alpha\beta}$.

We use natural system of units where the speed of light and the positron charge are equal to one.

Let us write down Maxwell's equations (1862) in relativistic form \cite{Vanderlinde}\footnote{We use Einstein convention about summation w.r.t. replicated indices.}
\begin{eqnarray}
\partial_\mu a_\nu-\partial_\nu a_\mu &=& f_{\mu\nu},\label{Maxw}\\
\partial_\mu f^{\mu\nu} &=& j^\nu,\nonumber
\end{eqnarray}
where $a_\mu\in\T_1$ is a potential of electromagnetic field,
$f_{\mu\nu}=-f_{\nu\mu}\in\T_2$ is a strength of electromagnetic field, and $j^\nu\in\T^1$ is a 4-vector of current. It follows from (\ref{Maxw}) that $\partial_\nu j^\nu=0$.
In the first equation in (\ref{Maxw}) there are two free (not contracted) indices $\mu,\nu$ and in the second equation
 there is one free index $\nu$. Therefore, the first equation is satisfied for all $\mu,\nu=0,1,2,3$
 and the second equation is satisfied for all $\nu=0,1,2,3$.

If we substitute $f_{\mu\nu}$ from the first equation of (\ref{Maxw}) into the second equation, then we get an equation of second order for electromagnetic potential
\begin{equation}
\partial_\mu\partial^\mu a^\nu-\partial^\nu(\partial_\mu a^\mu)=j^\nu.\label{Maxw:a}
\end{equation}
Systems of equations (\ref{Maxw}) and (\ref{Maxw:a}) are invariant w.r.t. a gauge transformation
\begin{eqnarray*}
a_\mu &\to& \acute a_\mu=a_\mu+\partial_\mu\sigma,\\
f_{\mu\nu} &\to& \acute f_{\mu\nu}=f_{\mu\nu},\\
j^\nu &\to& \acute j^\nu=j^\nu,
\end{eqnarray*}
where $\sigma=\sigma(x)$ is a twice differentiable function $\sigma : \R^{1,3}\to\R$.

\medskip

\noindent{\bf Pseudo-Euclidean space $\R^{p,q}$.} The relativistic form of Maxwell's equation gives us the possibility of considering these equations in arbitrary $n$ dimensional pseudo-Euclidean space $\R^{p,q}$, ($p,q$ are nonnegative integer numbers and $p+q=n$) with Cartesian coordinates $x^\mu$, $\mu=1,\ldots,n$ and with a metric tensor given by the diagonal $n\times n$- matrix
$$
\eta=\|\eta_{\mu\nu}\|=\|\eta^{\mu\nu}\|=\diag(1,\ldots,1,-1,\ldots,-1).
$$
with $p$ pieces of  $1$ and $q$ pieces of $-1$ on the diagonal.

In the sequel we consider Maxwell's equations (\ref{Maxw}) and other systems of equations in pseudo-Euclidean space $\R^{p,q}$.


\section{Proca equations}

In 1936 \cite{Proca} the Romanian physicist Alexandru Proca invented the following modification of the relativistic Maxwell equations:
\begin{eqnarray}
\partial_\mu a_\nu-\partial_\nu a_\mu &=& f_{\mu\nu},\label{Proca}\\
\partial_\mu f^{\mu\nu}+m^2 a^\nu &=& 0,\nonumber
\end{eqnarray}
where $m$ is a real constant (mass of a particle of spin $1$). From the equations (\ref{Proca}) for $m\neq 0$ it follows the condition (Lorentz gauge)
\begin{equation}
\partial_\mu a^\mu=0.\label{L:cond}
\end{equation}
Hence, it follows from (\ref{Maxw:a}) that the system of equations (\ref{Proca}) can be reduced to Klein-Gordon-Fock equation for each component of $a^\nu$
\begin{equation}
\partial_\mu\partial^\mu a^\nu+m^2 a^\nu=0.\label{RGF:a}
\end{equation}

Consider Maxwell's equations (\ref{Maxw}) with vector of current $j^\nu$ that satisfies the condition $\partial_\nu j^\nu=0$ and
the conditions
\begin{equation}
\partial_\mu\partial^\mu j^\nu+m^2 j^\nu=0.\label{RGF:j}
\end{equation}
In that case, Maxwell's equations have a solution
$$
a_\mu=-\frac{1}{m^2}j_\mu.
$$
This solution is not gauge invariant.

Consequently solutions of Proca equations can be considered as partial subclass of solutions of Maxwell's equations with right hand side (current $j^\nu$) that satisfies additional conditions (\ref{RGF:j}).

\section{Yang-Mills equations}

 Let $K$ be a semisimple Lie group; $L$ be the real Lie algebra of the Lie group $K$. A Lie algebra $L$ is a real vector space of dimension $N$ with basis $t^1,\ldots,t^N$. Multiplication of elements of $L$ is given by Lie bracket  $[A,B]=-[B,A]$, which satisfies Jacobi's identity. Multiplication of basis elements is given with the aid of real structural constants $c^{rs}_l=-c^{sr}_l$ ($r,s,l=1,\ldots,N$) of the Lie algebra $L$
\begin{equation}
[t^r,t^s]=c^{rs}_l t^l.\label{ttc}
\end{equation}
In this work we represent elements of the Lie algebra $L$ and the Lie group $K$ by square matrices of respective dimension or by elements of Clifford algebra $\cl(p,q)$. In both cases Lie bracket is given by commutator
$[A,B]=A B-B A$, where on right hand side we use matrix multiplication of matrices or Clifford multiplication of Clifford algebra elements.

By $L\T^a_b$ we denote a set of tensor fields of the pseudo-Euclidean space $\R^{p,q}$ of type $(a,b)$ and of rank $a+b$ with values in the Lie algebra $L$.

Consider the following equations in pseudo-Euclidean space $\R^{p,q}$:
\begin{eqnarray}
&&\partial_\mu A_\nu-\partial_\nu
A_\mu-\rho[A_\mu,A_\nu]=F_{\mu\nu},\label{YM}\\
&&\partial_\mu F^{\mu\nu}-\rho[A_\mu,F^{\mu\nu}]=J^\nu\nonumber
\end{eqnarray}
where  $A_\mu\in L\T_1$,
 $J^\nu\in
L\T^1$, $F_{\mu\nu}=-F_{\nu\mu}\in L\T_{2}$,  $\rho$ is a real constant (interaction constant). These equations are called {\em Yang-Mills equations} (system  of Yang-Mills equations).
One suggests that $A_\mu,F_{\mu\nu}$ are unknown and $J^\nu$ is known vector with values in Lie algebra $L$.
One says that equations (\ref{YM}) define {\em Yang-Mills field} $(A_\mu,F_{\mu\nu})$, where $A_\mu$ is {\em potential} and $F_{\mu\nu}$
is {\em strength} of Yang-Mills field.
A vector $J^\nu$ is called {\em non-Abelian current} (in the case of Abelian group $K$ vector  $J^\nu$ is called {\em
current}). {\sloppy

}

The components of the skew-symmetric tensor $F_{\mu\nu}$ from the first equation of (\ref{YM}) can be substituted into the second equation to get one equation of second order for the potential of Yang-Mills field
\begin{equation}
\partial_\mu(\partial^\mu A^\nu-\partial^\nu A^\mu-\rho[A^\mu,A^\nu])-\rho[A_\mu,\partial^\mu A^\nu-\partial^\nu A^\mu-\rho[A^\mu,A^\nu]]=J^\nu.\label{YM:A}
\end{equation}
Let us consider equation (\ref{YM}) from another point of view. Let $A_\mu\in L\T_1$ be arbitrary covector with values in $L$, which smoothly depends on $x\in\R^{p,q}$. By
$F_{\mu\nu}$ denote the expression
\begin{equation}
F_{\mu\nu}:=\partial_\mu A_\nu-\partial_\nu A_\mu-\rho[A_\mu,A_\nu],
\label{Fmunu}
\end{equation}
and by $J^\nu$ denote the expression
$$
J^\nu:=\partial_\mu F^{\mu\nu}-\rho[A_\mu,F^{\mu\nu}].
$$

Now we can consider the expression  $\partial_\nu
J^\nu-\rho[A_\nu,J^\nu]$ and, with the aid of simple calculations, we may verify that
\begin{equation}
\partial_\nu J^\nu-\rho[A_\nu,J^\nu]=0.
\label{nonabel:conslaw}
\end{equation}
This identity is called {\em non-Abelian conservation law} (in case of Abelian Lie group $K$ we have $\partial_\nu J^\nu=0$, i.e., divergence of the vector $J^\nu$ is equal to zero).

Therefore non-Abelian conservation law (\ref{nonabel:conslaw}) is a consequence of Yang-Mills equations (\ref{YM}).

Consider tensor fields $A_\mu,F_{\mu\nu},J^\nu$ that  satisfy Yang-Mills equations (\ref{YM}). Let us take a scalar field with values in Lie group $S=S(x)\in K$ and consider transformed tensor fields
\begin{eqnarray}
\acute A_\mu &=& S^{-1}A_\mu S-S^{-1}\partial_\mu S,\nonumber\\
\acute F_{\mu\nu} &=& S^{-1}F_{\mu\nu}S,\label{gauge:tr}\\
\acute J^{\nu} &=& S^{-1}J^{\nu}S.\nonumber
\end{eqnarray}
These tensor fields satisfy the same Yang-Mills equations
\begin{eqnarray*}
&&\partial_\mu\acute A_\nu-\partial_\nu
\acute A_\mu-\rho[\acute A_\mu,\acute A_\nu]=\acute F_{\mu\nu},\\
&&\partial_\mu\acute F^{\mu\nu}-\rho[\acute A_\mu,\acute F^{\mu\nu}]=
\acute J^\nu,
\end{eqnarray*}
i.e., equations (\ref{YM}) are invariant w.r.t. transformations (\ref{gauge:tr}). Transformation (\ref{gauge:tr})
is called {\em gauge transformation} (or {\em gauge symmetry}), and the Lie group $K$ is called {\em gauge group}
of Yang-Mills equations (\ref{YM}).

\medskip

\noindent{\bf Partial solutions of Yang-Mills equations.} During the last 60 years several classes of solutions of Yang-Mills equations were discovered. Namely, monopoles (Wu, Yang, 1968 \cite{WYa}), instantons (Belavin, Polyakov, Schwartz, Tyupkin, 1975 \cite{Bel}), merons (de Alfaro, Fubini, Furlan, 1976 \cite{deA}) and so on.\footnote{See review of Actor, 1979 \cite{Actor} and review of Zhdanov and Lagno, 2001 \cite{Zhdanov}.}


\section{Yang-Mills-Proca equations}

Let $K$ be semisimple Lie group; $L$ be the real Lie algebra of the Lie group $K$.
Consider equations in pseudo-Euclidean space $\R^{p,q}$
\begin{eqnarray}
&&\partial_\mu A_\nu-\partial_\nu
A_\mu-\rho[A_\mu,A_\nu]=F_{\mu\nu},\label{YMP}\\
&&\partial_\mu F^{\mu\nu}-\rho[A_\mu,F^{\mu\nu}]+m^2 A^\nu=0,\nonumber
\end{eqnarray}
where  $A_\mu\in L\T_1$,
 $F_{\mu\nu}=-F_{\nu\mu}\in L\T_{2}$;  $m,\rho$ are real constants and $[A,B]=-[B,A]$ is a Lie bracket. We call these equations {\em Yang-Mills-Proca equations (YMP)}.

Let us discuss some properties of this system of equations.

If $m\neq 0$ then YMP system of equations (\ref{YMP}) implies the identity (generalized Lorentz gauge)
\begin{equation}
\partial_\mu A^\mu=0.\label{div0}
\end{equation}
System of equation (\ref{YMP}) is not gauge invariant, but it is invariant w.r.t. a global (not dependent on $x\in\R^{p,q}$) transformation
$$
A_\mu\to\acute A_\mu=S^{-1}A_\mu S,\quad F_{\mu\nu}\to\acute F_{\mu\nu}=S^{-1}F_{\mu\nu}S,
$$
where $S$ is an element of a Lie group $K$ and $S$ is independent on $x$.

System of equations (\ref{YMP}) can be reduced to the following equation (system of equations) of second order for $A_\mu$:
\begin{equation}
\partial_\mu(\partial^\mu A^\nu-\partial^\nu A^\mu-\rho[A^\mu,A^\nu])-\rho[A_\mu,\partial^\mu A^\nu-\partial^\nu A^\mu-\rho[A^\mu,A^\nu]]+m^2 A^\nu=0.\label{YMP:A}
\end{equation}

For $m=0$ this equation coincides with the Yang-Mills equation (\ref{YM:A}) with the trivial right hand side ($J^\nu=0$).

Using the condition (\ref{div0}) and the formula
$$
\partial_\mu[A,B]=[\partial_\mu A, B] + [A, \partial_\mu B],
$$
the equation (\ref{YMP:A}) can be rewritten in the form
\begin{equation}
\partial_\mu\partial^\mu A^\nu - 2\rho[A^\mu,\partial_\mu A^\nu] + \rho[A_\mu,\partial^\nu A^\mu] +
\rho^2[A_\mu,[A^\mu,A^\nu]] + m^2A^\nu=0.\label{YMP:AA}
\end{equation}


\section{Constant solutions of Yang-Mills-Proca equations}\label{sect5}

We are looking for constant (not dependent on $x\in\R^{p,q}$) solutions $A_\mu\in L\T_1$ of Yang-Mills-Proca equations (\ref{YMP:A}) for fixed constant $m$. For this solutions
$$
\partial_\mu A_\nu=0.
$$
Therefore, the system of nonlinear differential equations (\ref{YMP:A}) reduces to the system of algebraic (cubic) equations
\begin{equation}
[A_\mu,[A^\mu,A^\nu]]=\lambda A^\nu,\label{A3}
\end{equation}
where  $\lambda=-m^2/{\rho^2}$. Constant solutions of Yang-Mills equations (when $\lambda=0$) are discussed in \cite{Sch} and \cite{Sch2}.

Let us write down components of vector $A^\mu$ with values in Lie algebra $L$ in the form of decomposition
w.r.t. a basis of the Lie algebra
$$
A^\mu=a^\mu_r t^r,
$$
where real coefficients $a^\mu_r$ ($\mu=1, \ldots, n$; $r=1, \ldots, N$) define $N$ vectors.
Substituting these decompositions into the equations (\ref{A3}) and using the relations (\ref{ttc}), we get $n N$ algebraic cubic equations for $n N$ unknown coefficients $a^\mu_r$. A resulting system of cubic equations contains a real parameter $\lambda$ and structure constants $c^{rs}_l$ of the Lie algebra\footnote{In case of Minkowski space $\R^{1,3}$ and (three dimensional) Lie algebra $\su(2)$ of special unitary Lie group $\SU(2)$ we have a system of 12 equations with 12 unknowns $a^\mu_r$ ($\mu=1, 2, 3, 4$; $r=1, 2, 3$).}.

We cannot give any standard method to solve a system of cubic equations (if we have one cubic equation with one unknown, then we can use Cardano's formula). Nevertheless, we have found (guess) several classes of solutions of the system of equations (\ref{A3})  (for $\lambda>0$, for $\lambda<0$, and for $\lambda=0$).
\medskip

\noindent{\bf Commuting solutions of the system of equations (\ref{A3}).} Any set of $n$ mutually commuting elements (matrices) $A_\mu$ of the Lie algebra $L$
$$
[A_\mu,A_\nu]=0
$$
is a solution of the system of equations (\ref{A3}) with $\lambda=0$ (i.e. $m=0$). Such solutions of Yang-Mills equations were considered by M.~Ikeda and Y.~Miyachi (1962, \cite{Ikeda}).

\medskip

\section{Anti-commuting solutions of Yang-Mills-Proca equations}\label{sect6}
Let us remind that we represent elements of the Lie group $L$ in the form of square matrices of some size or in the form of elements of Clifford algebra $\cl(p,q)$  and in both cases the Lie bracket is expressed by the commutator $[A,B]=A B-B A$. By ${\bf1}$ we denote the identity matrix of corresponding size or the  identity element of Clifford algebra.
\begin{thm}\label{theorem:1}
Consider the pseudo-Euclidian space $\R^{p,q}$ of dimension $n=p+q\geq2$. Let us take a parameter $\theta=1$ or $\theta=-1$.
If the Lie algebra $L$ contains $n$ elements $A_\mu$  such that
\begin{equation}
A_\mu A_\nu + A_\nu A_\mu=2\theta\eta_{\mu\nu}{\bf1},\label{AAAA}
\end{equation}
then these elements $A_\mu$ satisfy the system of equations (\ref{A3}) with $\lambda=4\theta(n-1)$.
\end{thm}

\proof Let $\theta=1$ and the Lie algebra $L$ contains $n$  elements (components of a covector) $A_\mu$ that satisfy relations (\ref{AAAA}). In other words, $A_\mu$ are such that
$$
A_\mu A_\nu= - A_\nu A_\mu,\quad \mu\neq\nu
$$
and
$$
(A_\mu)^2=\eta_{\mu\mu}{\bf1}.
$$
For such set of elements  $A_\mu$ we have relations
\begin{eqnarray}
&&A_\mu A^\nu A^\mu=A^\mu A^\nu A_\mu=(2-n)A^\nu,\label{theor:conv}\\
&&A^\mu A_\mu=n{\bf1},\nonumber
\end{eqnarray}
which follow from the theorem about generators contractions (see \cite{MarSh2008}, page 242).

Let us calculate the left hand side of the equation (\ref{A3}). Replacing Lie brackets by commutators and using formulas
 (\ref{theor:conv}) we get
\begin{eqnarray*}
[A_\mu,[A^\mu,A^\nu]] &=& A_\mu A^\mu A^\nu - A_\mu A^\nu A^\mu -A^\mu A^\nu A_\mu + A^\nu A^\mu A_\mu\\
&=& 2n A^\nu - 2 A_\mu A^\nu A^\mu\\
&=& 4(n-1)A^\nu.
\end{eqnarray*}
The proposition is proved. For $\theta=-1$ a proof is similar. $\blacksquare$

The set of elements $A_\mu$ (\ref{AAAA}) generates Cliford algebra of dimension $n$ or, in some cases, of dimension $n-1$ (see, for more details, Section \ref{Clif}). For example, in the case of real Clifford algebra of odd dimension $n=p+q$ and signatures $p-q=1\mod 4$, we have such solution (\ref{AAAA}) that elements $A_\mu$ are dependent. For example, if $n=3$, then for each of the signatures $(2,1)$, $(1,2)$, $(0,3)$  there exists a solution to the system of equations (\ref{AAAA}) such that $A_1,A_2,A_3$ are arbitrary variables that satisfy the condition  $\tr(A_1 A_2 A_3)=0$. But for signatures $(2,1)$ and
$(0,3)$ there exists an additional solution to the system of equations (\ref{AAAA}) of the form $A_1,A_2,A_3$, where $A_3=A_1 A_2$.
In this case $\tr(A_1 A_2 A_3)\neq0$.\label{dop}

\begin{thm}\label{theorem:2}
Consider the pseudo-Euclidean space $\R^{p,q}$ of dimension $n=p+q\geq2$. Let us take a parameter $\theta=1$ or $\theta=-1$.
Suppose that a set of $n$ elements (covector components) $A_\mu$ satisfy identities (\ref{AAAA}) and the Lie algebra $L$ contains a set of $n$ elements $\acute A_\mu$  such that this set is obtained from the set $A_\mu$ by taking  $r$ ($1\leq r\leq n-2$)  elements of this set equal to zero.
Then the set $\acute A_\mu$ satisfies the system of equations (\ref{A3}) with $\lambda=4\theta(\acute n-1)$, where $\acute n=n-r\geq2$.
\end{thm}
\proof A proof follows from the proof of Theorem \ref{theorem:1}. $\blacksquare$

\medskip

\noindent{\bf Multiplication of a solution by a constant.} Suppose that a covector  $A_\mu\in L\T_1$ is independent of $x$ and satisfies the system of equations (\ref{A3}), and $\kappa\in\R$ is a nonzero constant; then the covector
$\check A_\mu= \kappa A_\mu$ also satisfies the system of equations (\ref{A3}) but with the parameter
$$
\check\lambda = \kappa^2\lambda.
$$

\medskip

Let us summarize our reasoning. Suppose that a pair of nonnegative integer numbers $(p,q)$ defines a signature of the  pseudo-Euclidean space $\R^{p,q}$ of dimension $n=p+q\geq2$; then the number of pairs $(\acute p,\acute q)$ that satisfy the conditions
$$
\acute p\leq p,\quad \acute q\leq q,\quad \acute p+\acute q\geq 2
$$
are equal to $(p+1)(q+1)-3$. For ``appropriate'' Lie algebras $L$, any of these pair $(\acute p,\acute q)$ is connected to the pair of constant solutions $A_1,\ldots,A_n$ (for $\theta=\pm1$) of the system of equations (\ref{A3}) in the pseudo-Euclidean space $\R^{p,q}$ with constant $\lambda=4\theta(\acute p+\acute q-1)$.
These solutions are defined up to multiplication by a real nonzero constant
$\kappa$ (in this case the constant $\lambda$ is multipied by $\kappa^2$).

\medskip

\noindent{\bf What are ``appropriate'' Lie algebras $L$.} ``Appropriate'' Lie algebras $L$ must contain a subalgebra that is isomorphic to Clifford algebra $\cl^\R(p,q)$. Otherwise a number of considered constant anti-commutative solutions of Yang-Mills-Proca system of equations is decreased.

\medskip

\noindent{\bf Constant solutions of Yang-Mills-Proca system of equations in Minkowski space.}
As an example, consider constant solutions of Yang-Mills-Proca system of equations in Minkowski space $\R^{1,3}$ with Cartesian coordinates $x^\mu$, $\mu=0,1,2,3$ and with the diagonal metric tensor (\ref{eta13}). We need four vectors (tetrad) $y^\mu_a$, $\mu=0,1,2,3$, $a=0,1,2,3$, which are numbered by Latin index $a$ and satisfy relations
\begin{equation}
y^\mu_a y^\nu_b\eta^{ab}=\eta^{\mu\nu}.\label{tetrada}
\end{equation}
By Theorem \ref{theorem:1} we must take a covector $A_\mu$ with values in some real Lie algebra $L$ and components of this covector satisfy relations (\ref{AAAA}) with $\theta=1$, or $\theta=-1$. From the theory of Dirac equations we know that the set of four matrices  $\gamma^a$, $a=0,1,2,3$ in Dirac representation
\begin{eqnarray}
\gamma^0&=&\begin{pmatrix}1 &0 &0 & 0\cr
                  0 &1 & 0&0 \cr
                  0 &0 &-1&0 \cr
                  0 &0 &0 &-1\end{pmatrix},\quad
\gamma^1=\begin{pmatrix}0 &0 &0 & 1\cr
                  0 &0 & 1&0 \cr
                  0 &-1 &0 &0 \cr
                  -1 &0 &0 &0\end{pmatrix},
\label{gamma:matrices}
\\
\gamma^2&=&\begin{pmatrix}0 &0 &0 & -i\cr
                  0 &0 & i&0 \cr
                  0 & i&0 &0 \cr
                  -i &0 &0 &0\end{pmatrix},\quad
\gamma^3=\begin{pmatrix}0 &0 & 1& 0\cr
                  0 &0 & 0&-1 \cr
                  -1 &0 &0 &0 \cr
                  0 &1&0 &0\end{pmatrix}.
\nonumber
\end{eqnarray}
satisfies relations (\ref{AAAA}) with $\theta=1$ and the matrices $i\gamma^a$ satisfy conditions (\ref{AAAA}) with $\theta=-1$. The Hermitian conjugated matrices satisfy conditions
\begin{equation}
(\gamma^a)^\dagger=\gamma^0\gamma^a\gamma^0,\quad (i\gamma^a)^\dagger=-\gamma^0i\gamma^a\gamma^0.\label{gamma:dagger}
\end{equation}
Let us remind definitions of Lie group of special pseudo-unitary matrices $\SU(2,2)$ and its real Lie algebra $\su(2,2)$
\begin{eqnarray*}
\SU(2,2) &=& \{S\in\Mat(4,\C) : S^\dagger\beta S=\beta,\ \det\,S=1\},\\
\su(2,2) &=& \{s\in\Mat(4,\C) : \beta s^\dagger\beta=-s,\ \tr\,s=0\},
\end{eqnarray*}
where $\beta=\diag(1,1,-1,-1)$. From these definitions and from formulas (\ref{gamma:dagger}) we see that $i\gamma^a\in\su(2,2)$ ($\beta=\gamma^0$).

Whence if we take a Lie algebra $L=\su(2,2)$ then the following vector with values in $L$
\begin{equation}
A^\mu=\kappa y^\mu_a\gamma^a\label{A:def}
\end{equation}
satisfies conditions (\ref{AAAA}) for $\theta=-1$ and by Theorem \ref{theorem:1} this vector is a solution to the system of equations (\ref{A3}) with constant $\lambda=-12\kappa^2$, ($\kappa$ is real parameter). So, for the real Lie algebra  $L=\su(2,2)$, we get a constant solution of the Yang-Mills-Proca system of equations (\ref{YMP}) in Minkowski space $\R^{1,3}$ with real constants that are connected by the relation
$$
\frac{m^2}{\rho^2}=12\kappa^2.
$$
\medskip

\noindent{\bf Constant solutions of Yang-Mills-Proca system of equations in Euclidean space $\R^3$.}
As a second example let us consider constant solutions of Yang-Mills-Proca system of equations in Euclidean space $\R^{3}$ with Cartesian coordinates $x^\mu$, $\mu=1,2,3$ and with the diagonal metric tensor given by $3\times3$ identity matrix $\eta=\diag(1,1,1)$. We need three independent of $x$ vectors $y^\mu_a$, $\mu=1,2,3$, $a=1,2,3$, which are numbered by Latin index $a$ and satisfy relations
\begin{equation}
y^\mu_a y^\nu_b\eta^{ab}=\eta^{\mu\nu}.\label{tetrada}
\end{equation}
Let us remind definitions of Lie group of special unitary matrices $\SU(2)$ and its real Lie algebra $\su(2)$
\begin{eqnarray*}
\SU(2) &=& \{S\in\Mat(2,\C) : S^\dagger =S^{-1},\ \det\,S=1\},\\
\su(2) &=& \{s\in\Mat(2,\C) : s^\dagger=-s,\ \tr\,s=0\}.
\end{eqnarray*}
Let $\tau^a$ be the Pauli matrices multiplied by imaginary unit $i$
\begin{equation}
\tau^1=i\begin{pmatrix}0 &1\cr
                       1 & 0\end{pmatrix},\quad
\tau^2=i\begin{pmatrix}0 & -i\cr
                       i & 0\end{pmatrix},\quad
\tau^3=i\begin{pmatrix}1 & 0\cr
                       0 & -1\end{pmatrix},
\end{equation}
We see that $\tau^a\in\su(2)$. And these matrices satisfy conditions (\ref{A3}) with $\theta=-1$.
Therefore, if we take a Lie algebra $L=\su(2)$, then the following vector with values in $L$
\begin{equation}
A^\mu=\kappa y^\mu_a\tau^a\label{A:tau}
\end{equation}
satisfies conditions (\ref{AAAA}) with $\theta=-1$ and by Theorem \ref{theorem:1} this vector is a solution to the system of equations (\ref{A3}) with constant $\lambda=-8\kappa^2$. So, in the Euclidean space $\R^3$  we get a constant solution of Yang-Mills-Proca system of equations with Lie algebra $L=\su(2)$ and with real constants that are connected by the relation
$$
\frac{m^2}{\rho^2}=8\kappa^2.
$$
Note that this example deals with a class of additional solutions (with $\tr\,A_1A_2A_3\neq0$) which were considered after the proof of theorem \ref{theorem:1}.


\section{Solutions of Yang-Mills equations in the form of perturbation theory series}

Consider the pseudo-Euclidean space $\R^{p,q}$ of dimension $n=p+q\geq2$ with Cartesian coordinates $x^\mu$ and with the metric tensor $\eta_{\mu\nu}$. Let $\gamma^\mu$ be constant (independent of $x$) vector field (with values in matrix algebra or in Clifford algebra) such that components $\gamma^\mu$ satisfy the relations $\gamma^\mu\gamma^\nu+\gamma^\nu\gamma^\mu=2\theta\eta^{\mu\nu}{\bf1}$,
where the parameter $\theta=1$ or $\theta=-1$. A real Lie algebra $L$ is such that $\gamma^\mu\in L\T^1$. Now we consider  the system of Yang-Mills equations (\ref{YM:A}) with the Lie algebra $L$, with the parameter $\rho=1$, and with right hand side
\begin{equation}
J^\nu=4\theta(n-1)\gamma^\nu. \label{JJ}
\end{equation}
By Theorem \ref{theorem:1} this system of Yang-Mills equations, in particular, has constant solution $A_\mu=\gamma_\mu$ and $[\gamma_\mu,[\gamma^\mu,\gamma^\nu]]=4\theta(n-1)\gamma^\nu$.

Our aim is to consider solutions of Yang-Mills equations in the form of perturbation theory series near the constant solution $A_\mu=\gamma_\mu$. We take a small parameter  $\varepsilon<1$ and substitute the expression
$$
A_\mu= \sum_{k=0}^\infty \varepsilon^k \st{k}{A_\mu}
$$
into the left hand side of equation (\ref{YM:A}). Let us write the result in the form of power series w.r.t. $\varepsilon$
\begin{eqnarray*}
\partial_\mu(\partial^\mu A^\nu-\partial^\nu A^\mu-[A^\mu,A^\nu])-[A_\mu,\partial^\mu A^\nu-\partial^\nu A^\mu-[A^\mu,A^\nu]]=\sum_{k=0}^\infty \varepsilon^k Q_k^\nu=4\theta(n-1)\gamma^\nu,
\end{eqnarray*}
where
\begin{eqnarray}
Q_k^\nu &=& \partial_\mu\partial^\mu\st{k}{A^\nu} - \partial^\nu\partial_\mu\st{k}{A^\mu}-\sum_{l=0}^k ([\partial_\mu\st{l}{A^\mu},\st{k-l}{A^\nu}]+[\st{l}{A^\mu},\partial_\mu\st{k-l}{A^\nu}])\label{Qk}\\
&& - \sum_{s=0}^k[\st{k-s}{A_\mu},
\partial^\mu\st{s}{A^\nu}-\partial^\nu\st{s}{A^\mu} - \sum_{r=0}^s[\st{r}{A^\mu},\st{s-r}{A^\nu}]].\nonumber
\end{eqnarray}
For every integer $k\geq0$, the components of vector $Q_k^\nu$ depend on
$\st{0}{A_\mu},\ldots,\st{k}{A_\mu}$. Therefore some approximate solutions of Yang-Mills equations (\ref{YM:A}) with the right hand side (\ref{JJ}) can be found with the aid of the following procedure. Let us take $\st{0}{A_\mu}=\gamma_\mu$. Then we get
$$
Q_0^\nu=[\gamma_\mu,[\gamma^\mu,\gamma^\nu]]=4\theta(n-1)\gamma^\nu.
$$
Substitute $\st{0}{A_\mu}=\gamma_\mu$ into the expression (\ref{Qk})
for $k=1$ and take
\begin{equation}
Q_1^\nu=0.\label{Q1}
\end{equation}
As a result, we get a system of linear partial differential equations with constant coefficients and with variables $\st{1}{A_\mu}$. Let us take any solution of this system of equations (for example, a plane wave solution) and let us substitute this solution (together with $\st{0}{A_\mu}=\gamma_\mu$) into the expressions
$$
Q_2^\nu=0.
$$
Now we get a system of linear partial differential equations with variable coefficients (dependent on $x\in\R^{p,q}$) for variables $\st{2}{A_\mu}$. Again we take any solution of this system of equations and substitute this solution into the expression $Q_3^\nu=0$. Continuing  this procedure, we get $\st{k}{A_\mu}$ for any integer $k\geq0$. So we get an approximate solution (up to terms of order $\varepsilon^{k}$) of Yang-Mills equations with the right hand side (\ref{JJ}).

Let us summarize our reasoning.
  If we look for approximate solutions of Yang-Mills system of equations with the right hand side (\ref{JJ}) near the  constant solution $A_\mu=\gamma_\mu$, then we arrive at one linear system of partial differential equations with constant coefficients (for $\st{1}{A_\mu}$) and at sequence of linear systems of partial differential equations with variable coefficients
 (for $\st{2}{A_\mu},\ldots,\st{k}{A_\mu}$).

Let us consider in more details the system of equations  (\ref{Q1}) for $\st{1}{A_\mu}\equiv B_\mu$
\begin{eqnarray}
Q^\nu_1&\equiv&\partial_\mu\partial^\mu B^\nu -
\partial^\nu\partial_\mu B^\mu + [\gamma^\nu,\partial_\mu B^\mu]
-2[\gamma^\mu,\partial_\mu B^\nu] + [\gamma_\mu,\partial^\nu B^\mu]\label{linearized}\\
&& + [\gamma_\mu,[\gamma^\mu,B^\nu]] + [\gamma_\mu,[B^\mu,\gamma^\nu]] +
[B_\mu,[\gamma^\mu,\gamma^\nu]]=0,\nonumber
\end{eqnarray}
which are the linearization of Yang-Mills system of equations (\ref{YM:A})
(with parameter $\rho=1$  and with the right hand side $J^\nu=4\theta(n-1)\gamma^\nu$) near the constant solution $A_\mu=\gamma_\mu$.
For this system of equations (\ref{linearized}) one can easily find a class of simple solutions. Namely, let  $B_\mu$ be a vector such that every component of this vector commute with all $\gamma^\nu$. For even $n$ we have  $B_\mu=b_\mu{\bf1}$, where
$b_\mu=b_\mu(x)$  is a covector that satisfies Maxwell's equations with zero right hand side
$$
\partial_\mu\partial^\mu b^\nu - \partial^\nu\partial_\mu b^\mu=0.
$$
For odd $n$ we have $B_\mu=b_\mu{\bf1}+\hat b_\mu
\gamma^1\ldots\gamma^n$, where covectors $b_\mu=b_\mu(x)$, $\hat
b_\mu=\hat b_\mu(x)$ satisfy Maxwell's equations with zero right hand side ($j^\nu=0$). Evidently such $B_\mu$ satisfy the equations
(\ref{linearized}).

\section{Yang-Mills-Proca equations in Clifford algebra}\label{Clif}

Let us recall the basic notation. We consider real $\cl^\R(p,q)$ \cite{Clifford} and complexified $\cl^\C(p,q)=\C\otimes\cl^\R(p,q)$ \cite{Lounesto} Clifford algebra, $p+q=n$. In the general case, we write $\cl^\F(p,q)$, where $\F=\R, \C$. The construction of real and complexified Clifford algebras is discussed in details in \cite{Lounesto}, \cite{MarShir:book} and \cite{MarSh2008}.

Let $e$ be the identity element and $e^a$, $a=1,\ldots,n$ \cite{Benn:Tucker} be generators of the Clifford algebra $\cl^\R(p,q)$. Generators satisfy conditions $e^a e^b+e^b e^a=2\eta^{ab}e$, where $\eta=||\eta^{ab}||$ is the diagonal matrix with $p$ pieces of $+1$ and $q$ pieces of $-1$ on the diagonal. Elements $e^{a_1\ldots a_k}=e^{a_1}\cdots e^{a_k}$, $a_1<\cdots<a_k$, $k=1,\ldots,n$, together with the identity element $e$, form the basis of Clifford algebra.

We denote by $\cl^\R_k(p,q)$ the vector space spanned by the basis elements $e^{a_1\ldots a_k}$. Elements of $\cl^\R_k(p,q)$ are said to be elements of grade $k$. We have $\cl^\R(p,q)=\bigoplus_{k=0}^{n}\cl^\R_k(p,q).$

Clifford algebra can be considered as a Lie algebra with respect to the commutator $[U,V]=UV-VU$, $U,V\in\cl^\F(p,q)$. It is well-known that the following set is a center of Clifford algebra
$${\rm Cen}(\cl^\F(p,q))=\left\lbrace\begin{array}{ll}
\cl^\F_0(p,q), & \mbox{if $n$ is even};\\
\cl^\F_0(p,q)\oplus \cl^\F_n(p,q) & \mbox{if $n$ is odd.}
\end{array}
\right.
$$
The following set
$$\cl^\F_\circledS(p,q)=\cl^\F(p,q)\setminus{\rm Cen}(\cl^\F(p,q))$$
is a Lie subalgebra of Clifford algebra (see \cite{prim}).

Now we are looking for the constant solutions of Yang-Mills-Proca equations in the case of Lie algebra $L=\cl^\F_\circledS(p,q)$. We have
\begin{eqnarray}
[A_\mu,[A^\mu, A^\nu]]=\lambda A^\nu,\label{YMPs}
\end{eqnarray}
where $A^\mu\in L=\cl^\F_\circledS(p,q)$.

We have
$$[A_\mu,[A^\mu, A^\nu]]=A_\mu A^\mu A^\nu - A_\mu A^\nu A^\mu-A^\mu A^\nu A_\mu+A^\nu A^\mu A_\mu=\{A^\nu, A^\mu A_\mu\}-2 A_\mu A^\nu A^\mu,$$
where $\{U,V\}=UV+VU$ is anticommutator. So, equations (\ref{YMPs}) can be rewritten in the following form
\begin{eqnarray}
\{A^\nu, A^\mu A_\mu\}-2 A_\mu A^\nu A^\mu=\lambda A^\nu.\label{YMPs2}
\end{eqnarray}

It is easy to see (see also section \ref{sect6}), that there is the following class of solutions of these equations:
\begin{eqnarray}
(A^\mu)^2=\frac{\lambda\eta^{\mu\mu}e}{4(n-1)},\quad \mu=1, 2, \ldots, n;\qquad \{A^\mu, A^\nu\}=0,\quad \mu\neq \nu.\label{antics}
\end{eqnarray}
Really, we have
$$\{e^\nu, e^\mu e_\mu\}-2 e_\mu e^\nu e^\mu=2n e^\nu -2(2-n)e^\mu=(2n-4+2n)e^\nu=4(n-1)e^\nu,$$
because of the property $e_a e^b e^a=(2-n)e^b$ of Clifford algebra generators (see \cite{MarSh2008}).

Note, that after normalization such elements $A^\mu$ (\ref{antics}) will be generators: 1) of Clifford algebra $\cl^\F(p,q)$ or $\cl^\F(q,p)$, $p+q=n$ (in the case of real Clifford algebra $\F=\R$ there are 2 cases of signatures: $(p,q)$ and $(q,p)$; complex Clifford algebra $\F=\C$ does not depend on the signature); 2) of Clifford algebra of smaller dimension $n-1$ (for $p-q=1\mod 4$ in the case of real Clifford algebra and for $p-q=1, 3\mod 4$ in the case of complex Clifford algebra, see \cite{Porteous} and \cite{ShirokovPauli}); 3) of Grassmann algebra (for $\lambda=0$, see Section \ref{sectGr}).

Also, there are proportional (they commute, see Section \ref{sect5}) solutions $A_\mu$ of equations (\ref{YMPs2}) with $\lambda=0$ because of the form (\ref{YMPs}). Also, there are such solutions $A_\mu$ that some of them equal zero and the remaining ones generate a basis of Clifford algebra of smaller dimension (see Theorem \ref{theorem:2}).

Let us consider some examples in the cases of small dimensions $n=2, 3$.

\textbf{$n=2$.} In this case we have $\cl^\F_\circledS(p,q)=\cl^\F_1(p,q)\oplus \cl^\F_2(p,q)$. From (\ref{YMPs2}) we get
\begin{eqnarray}
(A^2)^2 A^1+A^1 (A^2)^2 - 2 A^2 A^1 A^2=\lambda \eta^{22}A^1,\nonumber\\
(A^1)^2 A^2+A^2 (A^1)^2 - 2 A^1 A^2 A^1=\lambda \eta^{11}A^2.\nonumber
\end{eqnarray}
Using $(A^1)^2$, $(A^2)^2$, $\{A^1, A^2\} \in \cl^\F_0(p,q)={\rm Cen}(\cl^\F(p,q))$ we obtain
\begin{eqnarray}
(A^2)^2 A^1 -  A^2 A^1 A^2=\lambda \frac{\eta^{22}}{2}A^1,\nonumber\\
(A^1)^2 A^2 -  A^1 A^2 A^1=\lambda \frac{\eta^{11}}{2}A^2.\nonumber
\end{eqnarray}
Further,
\begin{eqnarray}
2(A^2)^2 A^1 -  A^2 \{A^1, A^2\}=\lambda \frac{\eta^{22}}{2}A^1,\nonumber\\
2(A^1)^2 A^2 -  A^1 \{A^1, A^2\}=\lambda \frac{\eta^{11}}{2}A^2\nonumber
\end{eqnarray}
and
\begin{eqnarray}
2((A^2)^2 e-\lambda \frac{\eta^{22}}{4}e) A^1 - A^2 \{A^1, A^2\}=0,\label{systs}\\
2((A^1)^2 e- \lambda \frac{\eta^{11}}{4}e) A^2 -  A^1 \{A^1, A^2\}=0.\nonumber
\end{eqnarray}
It is easy to see that the following expressions are solution to this system of equations:
$$(A^1)^2=\frac{\lambda\eta^{11}}{4}e,\qquad (A^2)^2=\frac{\lambda\eta^{22}}{4}e,\qquad \{A_1, A_2\}=0.$$
If one of 4 scalar expressions in (\ref{systs}) does not equal to zero, then we obtain proportional solutions
$$A_1=\mu A_2,\qquad \mu=\frac{\{A^1,A^2\}}{2 (A^2)^2}=\frac{2 (A^1)^2}{\{A^1, A^2\}}$$
(or analogously $A_2=\mu A_1$), or one of $A_1$, $A_2$ equals to zero (we have $\lambda=0$ in these cases). So, we obtain all solutions of the system of equations (\ref{YMPs}) in the case $n=2$ for $L=\cl^\F_\circledS(p,q)$.

\textbf{$n=3$.} In this case we have $\cl^\F_\circledS(p,q)=\cl^\F_1(p,q)\oplus \cl^\F_2(p,q)$.
System of 3 equations (\ref{YMPs2}) for $A^1, A^2, A^3$ can be rewritten in the following form
\begin{eqnarray}
\eta^{22}((A^2)^2 A^1+A^1 (A^2)^2 - 2 A^2 A^1 A^2)+\eta^{33}((A^3)^2 A^1+A^1 (A^3)^2 - 2 A^3 A^1 A^3)=\lambda A^1,\nonumber\\
\eta^{33}((A^3)^2 A^2+A^2 (A^3)^2 - 2 A^3 A^2 A^3)+\eta^{11}((A^1)^2 A^2+A^2 (A^1)^2 - 2 A^1 A^2 A^1)=\lambda A^2,\nonumber\\
\eta^{11}((A^1)^2 A^3+A^3 (A^1)^2 - 2 A^1 A^3 A^1)+\eta^{22}((A^2)^2 A^3+A^3 (A^2)^2 - 2 A^2 A^3 A^2)=\lambda A^3.\nonumber
\end{eqnarray}
Using $(A^i)^2\in\cl^\F_0(p,q)\oplus \cl^\F_3(p,q)$ and $\{A_i, A_j\}\in\cl^\F_0(p,q)\oplus \cl^\F_3(p,q)={\rm Cen}(\cl^\F(p,q))$, we obtain
\begin{eqnarray}
\eta^{22}((A^2)^2 A^1-  A^2 A^1 A^2)+\eta^{33}((A^3)^2 A^1-  A^3 A^1 A^3)=\frac{\lambda}{2} A^1,\nonumber\\
\eta^{33}((A^3)^2 A^2-  A^3 A^2 A^3)+\eta^{11}((A^1)^2 A^2 -  A^1 A^2 A^1)=\frac{\lambda}{2} A^2,\nonumber\\
\eta^{11}((A^1)^2 A^3 -  A^1 A^3 A^1)+\eta^{22}((A^2)^2 A^3 -  A^2 A^3 A^2)=\frac{\lambda}{2} A^3,\nonumber
\end{eqnarray}
and
\begin{eqnarray}
\eta^{22}(2(A^2)^2 A^1- A^2 \{A^1, A^2\})+\eta^{33}(2(A^3)^2 A^1-  A^3 \{A^1, A^3\})=\frac{\lambda}{2} A^1,\nonumber\\
\eta^{33}(2(A^3)^2 A^2-  A^3 \{A^2, A^3\})+\eta^{11}(2(A^1)^2 A^2 - A^1 \{A^2 A^1\})=\frac{\lambda}{2} A^2,\nonumber\\
\eta^{11}(2(A^1)^2 A^3 - A^1 \{A^3, A^1\})+\eta^{22}(2(A^2)^2 A^3 - A^2 \{A^3, A^2\})=\frac{\lambda}{2} A^3,\nonumber
\end{eqnarray}
and
\begin{eqnarray}
A^1(2\eta^{22}(A^2)^2+2\eta^{33}(A^3)^2-\frac{\lambda}{2}e)+A^2(-\eta^{22}\{A^1, A^2\})+A^3(-\eta^{33}\{A^1, A^3\})=0,\nonumber\\
A^1(-\eta^{11}\{A^2, A^1\})+A^2(2\eta^{33}(A^3)^2+2\eta^{11}(A^1)^2-\frac{\lambda}{2}e)+A^3(-\eta^{33}\{A^2, A^3\})=0,\nonumber\\
A^1(-\eta^{11}\{A^3, A^1\})+A^2(-\eta^{22}\{A^3, A^2\})+A^3(2\eta^{11}(A^1)^2+2\eta^{22}(A^2)^2-\frac{\lambda}{2}e)=0.\nonumber
\end{eqnarray}
Elements in round brackets are elements of the center of Clifford algebra.
If they equal to zero, then we obtain the following solution of the system of equations:
$$\eta^{11}(A^1)^2=\eta^{22}(A^2)^2=\eta^{33}(A^3)^2=\frac{\lambda}{8}e,\qquad \{A_i, A_j\}=0.$$
To obtain other solutions we must consider all the remaining cases (if at least one of expressions in round brackets does not equal to zero). It is easy to see, that among solutions there will be proportional solutions $A_\mu$ with $\lambda=0$; commuting solutions with $\lambda=0$; solutions like: $A_1=0$ and 2 elements $A_2$, $A_3$ generate basis of Clifford algebra of dimension $n=2$, and similar others.

\section{Grassmann numbers as solutions of Yang-Mills-Proca equations in Clifford algebra}\label{sectGr}

Now we want to discuss one another class of solutions of Yang-Mills-Proca equations (\ref{YMPs}) with $\lambda=0$. It is easy to see that Grassmann numbers \cite{grass} are solutions of these equations. If we take Lie algebra $L=\cl^\C_\circledS(p,q)$ (let us consider only complex case), then we must realize Grassmann algebra as a subalgebra of Clifford algebra. We denote complexified Grassmann algebra of dimension $n$ by $\Lambda^\C(n)$. We can also consider degenerate Clifford algebra $\cl^\C(p,q,r)$ in more generale case.

We have the following well-known construction (see Clifford-Jordan-Wigner representation \cite{jw}, \cite{cjw}). Let us consider complex Clifford algebra $\cl^\C(n)=\cl^\C(n,0)$ of even dimension $n=p+q=2N$ (or odd dimension $n=2N+1$). With the use of generators $e^a$ we can construct the following elements
$$\theta^k=\frac{1}{2}(e^k+ie^{N+k}),\qquad k=1, \ldots, N,$$
$$\pi^k=\frac{1}{2}(e^k-i e^{N+k}),\qquad k=1, \ldots, N.$$
Note, that in the opposite way we have:
$$e^k=\theta^k+\pi^k,\qquad e^{k+N}=i(\theta^k-\pi^k).$$

It is easy to verify that these elements satisfy conditions
$$\theta^k \pi^l+\pi^l \theta^k=\delta^{kl},\qquad \theta^k\theta^l+\theta^l\theta^k=0,\qquad \pi^k \pi^l+\pi^l \pi^k=0.$$
So, we have 2 sets $\theta^k$ and $\pi^k$ of Grassmann numbers with some connections between each other.

Now let us consider degenerate Clifford algebras $\cl^\C(p,q,r)$ (see \cite{degen}, \cite{degen2}) with generators
$$e^1, \ldots, e^p,\quad \epsilon^1, \ldots, \epsilon^q, \quad \theta^1, \ldots, \theta^r,$$
where $(e^k)^2=1$, $(\epsilon^l)^2=-1$, $(\theta^m)^2=0$ for $k=1, \ldots, p$, $l=1, \ldots, q$, $m=1, \ldots, r$.

Jacobson radical (intersection of all maximal ideals) consists of elements
$$I=\sum_A a_A \theta^A+\sum_{A, B}b_{AB}e^A\theta^B+\sum_{A, B}c_{AB}\epsilon^A \theta^B+\sum_{A, B, C}d_{ABC}e^A \epsilon^B \theta^C,$$
and it is nilpotent. Algebra $\cl^\C(p,q,r)$ is not semi-simple. But it is well known that we can realize it in matrix algebra in the following way. Consider $\psi: \cl^\C(p,q,r)\to \cl^\C(p+r, q+r)$
\begin{eqnarray}
e^k&\to& e^k,\qquad k=1,\ldots, p,\nonumber\\
\epsilon^l &\to& \epsilon^l,\qquad l=1, \ldots, q,\nonumber\\
\theta^m&\to& e^{p+m}+\epsilon^{q+m},\qquad m=1,\ldots, r.\nonumber
\end{eqnarray}

For example, $\cl^\C(0,0,2)=\Lambda^\C(2)\to \cl^\C(2,2)$, with $\theta^1\to e^1+\epsilon^1$ and $\theta^2\to e^2+\epsilon^2$. We can consider standard matrix representation of square complex matrices of order 4. Degenerate Clifford algebra is a subalgebra of this algebra of matrices.

So, in Clifford algebra $\cl^\C(p,q)$ we can realize the following algebras
$$\cl^\C(p-1,q-1,1),\qquad \cl^\C(p-2,q-2,2),\qquad \ldots,\qquad \cl^\C(p-m,q-m,m),$$
where $m=min(p,q)$.

Grassmann algebra of even dimension $n$ can be represented using square complex matrices of order $2^n$ (it is isomorphic to a subalgebra of the algebra of such matrices), whenever complex Clifford algebra is isomorphic to an algebra of square complex matrices of order $2^{\frac{n}{2}}$. That is why we can always realize $\frac{n}{2}$ Grassmann numbers in Clifford algebra $\cl^\C(p,q)$, $p+q=n$.

Let us give one example. In the case of signature $(p,q)=(1,3)$ we have the following solution of (\ref{YMPs}):
$$A_1=T^{-1}a(ie^{23}-e^{13})T,\qquad A_2=T^{-1}b(e^{03}-e^3)T,\qquad A_3=0,\qquad A_4=0;\qquad \lambda=0$$
for any invertible element $T\in\cl^\C(p,q)$ and $a, b\in\C$.
Really, these Clifford algebra elements satisfy conditions of Grassmann algebra:
$A_1^2=A_2^2=0,\quad A_1 A_2=-A_2 A_1$.
In this case we can use the following 2 matrices:
\begin{eqnarray}
A_1=\begin{pmatrix}0 &0 &0 & 0\cr
                  1 &0 & 0&0 \cr
                  0 &0 &0&0 \cr
                  0 &0 &1 &0\end{pmatrix},\qquad
A_2=\begin{pmatrix}0 &0 &0 & 0\cr
                  0 &0 & 0&0 \cr
                  1 &0 &0 &0 \cr
                  0 &-1 &0 &0\end{pmatrix}.
\nonumber
\end{eqnarray}

\end{document}